\documentclass[12pt]{article}
\usepackage{cite}
\usepackage[T2A]{fontenc}
\usepackage[cp1251]{inputenc}
\usepackage{graphicx}
\usepackage{amsmath}
\title{Multifractal analysis of stress time series during ultrathin lubricant film melting}
\author{A.V.~Khomenko, I.A.~Lyashenko, V.N.~Borisyuk\\
Sumy State University, 40007, Sumy, Ukraine 
\\ khom@mss.sumdu.edu.ua, nabla04@ukr.net, el\_vado@ukr.net}
\date{\ }
\begin{document}
\maketitle
\thispagestyle{empty}
\def\abstractname{\ }
\begin{abstract}
\vspace{-2cm}
Melting of an ultrathin lubricant film confined between two atomically flat surfaces is we studied using the rheological model for viscoelastic matter approximation. Phase diagram with domains, corresponding to sliding, dry, and two types of $stick-slip$ friction regimes has been built taking into account additive noises of stress, strain, and temperature of the lubricant.
The stress time series have been obtained for all regimes of friction using the Stratonovich interpretation. It has been shown that self-similar regime of lubricant melting is observed
when intensity of temperature noise is much larger than intensities of strain and stress noises. This regime is defined by homogenous distribution, at which characteristic stress scale is absent. We study stress time series obtained for all friction regimes using multifractal detrended fluctuation analysis. It has been shown that multifractality of
these series is caused by different correlations that are present in the system and also by a power-law distribution. Since the power-law
distribution is related to small stresses, this case corresponds to self-similar solid-like lubricant.

\noindent {\bf Keywords:} White noise; time series; Fokker-Planck and Langevin equations; correlations; melting; stick-slip friction.

\noindent {\bf PACS:} 05.10.Gg, 05.30.Pr, 05.45.-a, 05.45.Tp, 07.05.Tp, 62.20.Fe, 62.20.Qp, 64.60.-i, 68.60.-p
\end{abstract}

\section{Introduction}

Problem of sliding friction is of great interest due to its applied engineering significance \cite{1lit}. Atomically flat surfaces separated by ultrathin layer of lubricant are under active investigation recently. These systems exhibit anomalous behavior, consisting in existence of several
kinetic regimes of friction. Transitions between the regimes are interpreted as a phase transitions \cite{land_inst}.
The liquid lubricant shows properties of solids \cite{Yosh}. Distinctive peculiarity of the systems is inherent in dry friction interrupted motion
($stick-slip$ regime) \cite{2lit,3lit,Carlson,Aranson,Filippov}. Denoted regime is observed for lubricant
thickness equal or less than three molecular layers, and is explained by periodical solidification due to walls pressing. Sheared
lubricant melts when shear stresses $\sigma$ are larger than the critical value $\sigma_c$ (yield point) owing to ``shear melting'' effect. The increased interest to such systems has motivated appearance of several models. Deterministic model \cite{Carlson}, thermodynamic model \cite{Popov}, and rheological model \cite{liq} were developed to describe above mentioned properties. Investigations are also based on molecular dynamics methods \cite{Braun}. The influence of additive non-correlated noises of basic parameters \cite{pla,stoh} and correlated fluctuations of temperature \cite{uhl_fnl} on lubricant melting has been investigated within the framework of rheological model \cite{liq}. Reasons for hysteresis behavior \cite{FTT,PhysLettA,JPS} and melting due to dissipative heating of friction surfaces \cite{dissipative} have been also considered. Systems with different viscosity dependence on temperature are also analyzed \cite{CondMat}.

We suppose that with the increase in stress $\sigma$ the lubricant melts, since the velocity of moving
surfaces also increases according to the relationship \cite{wear,dissipative}:
\begin{equation}
V = \sigma\frac{h}{\eta_{eff}},
\label{V_block}
\end{equation}
where $h$ is the thickness of lubricant or distance between friction surfaces, $\eta_{eff}$ is the effective viscosity, being measured
experimentally \cite{wear}.

The present work is devoted to time dependencies investigation of stresses in a self-similar regime of lubricant melting, caused by temperature
fluctuations. This regime was found in \cite{stoh} based on the method described in \cite{physa_soc, TorKhar}.

\section{Dynamic phase diagram}

In previous works \cite{liq,pla,stoh,dissipative} we treated a viscoelastic medium with a non-zero thermal conductivity
using the rheological model. The system of kinetic equations was also derived describing mutually coordinated evolution
of shear stress $\sigma$ and strain $\varepsilon$, and temperature $T$ in ultrathin lubricant film during
friction between atomically flat solid surfaces. We used the measure units
\begin{eqnarray}
\sigma_{s}{=}\left( {\rho c_{\upsilon}\eta_{0}T_{c} \over \tau_{T}}\right)^{1/2}, \quad
\varepsilon_{s}{=}{\frac{\sigma_{s}}{G_{0}}},\quad
T_{c} \label{1aa} \end{eqnarray} for variables
$\sigma$, $\varepsilon $, $T$, respectively, where $\rho$ is the lubricant density,
$c_v$ is the specific heat capacity, $T_{c}$ is the critical temperature,
$\eta_{0} \equiv \eta$ at $T=2T_{c}$ is the characteristic
value of shear viscosity $\eta$, $\tau_{T}\equiv\rho h^2 c_{\upsilon}/\kappa$ and $h$ are the time of heat conduction and
thickness of lubricant, $\kappa$ is the thermal conductivity,
$\tau_{\varepsilon}\sim 10^{-12}$~s is the relaxation time of matter
strain, $G_{0}\equiv \eta _{0}/\tau _{\varepsilon}$ is the characteristic value of shear modulus. Let us write the equations:
\begin{eqnarray}
&&\tau _{\sigma}\dot{\sigma}=-\sigma^a + g\varepsilon+\sqrt{I_\sigma}\xi_1(t), \label{eq2} \\
&&\tau_\varepsilon\dot\varepsilon=-\varepsilon + (T-1)\sigma^a+\sqrt{I_\varepsilon}\xi_2(t),\label{eq3} \\
&&\tau_T\dot T=(T_e-T) - \sigma^a\varepsilon +\sigma^{2a}+\sqrt{I_T}\xi_3(t). \label{eq4}
\end{eqnarray}
Here the stress relaxation time $\tau_{\sigma}$, the temperature $T_{e}$ of atomically flat solid friction surfaces, and the
constant $g=G/G_{0}<1$ are introduced, where $G$ is the lubricant shear modulus. Quantities $I_\sigma$, $I_\varepsilon$, and $I_T$ are the
intensities of stress, strain, and temperature noises, respectively.  Substitution of $\partial \varepsilon / \partial t$ instead of $\varepsilon / \tau_{\sigma}$ in Eq.~(\ref{eq2}) reduces it to a
Maxwell-type equation for a viscoelastic matter, which
is widely used in the theory of boundary friction \cite{1lit}.
The relaxation behavior of a viscoelastic lubricant during
friction is also described by the Kelvin-Voigt equation~(\ref{eq3})
\cite{liq,voigt}. It takes into account the dependence of the shear
viscosity on the dimensionless temperature $\eta = \eta_{0}/(T-1)$.
Equation~(\ref{eq4}) is heat conduction expression describing heat transfer from friction surfaces to the
layer of lubricant, the dissipative heating of the stress-induced viscous flow,
and a heat source due to the reversible mechanocaloric effect.
Equations ~(\ref{eq2}) -- (\ref{eq4}) formally coincide with the
Lorenz synergetic system \cite{Lorenz,13}, where the shear stress
acts as the order parameter, the conjugate field is reduced to the shear strain, and temperature is the control
parameter. When $\sigma = 0$ the lubricant is solid-like, situation with $\sigma \ne 0$ corresponds to its liquid-like
state \cite{liq,stoh,uhl_fnl,FTT,PhysLettA,JPS,dissipative,CondMat}.

In equations (\ref{eq2}) -- (\ref{eq4}) $0{<}a{<}1$ is the fractional exponent. The function $\xi_i(t)$ is $\delta$-correlated Gaussian source
(white noise). Its moments are defined as\footnote{Here multiplier $2$ is chosen for simplification of the
corresponding Fokker -- Planck equation (FPE).}:
\begin{equation}
\langle\xi_i(t)\rangle = 0,\quad
\langle\xi_i(t)\xi_j(t')\rangle = 2\delta_{ij}\delta(t-t').
\label{xi_corr}
\end{equation}
Experimental data for organic lubricant \cite{Yosh} show
that relaxation time of the stress
$\tau_{\sigma}$ at normal pressure is $\sim 10^{-10}$~s, and it increases
by several orders of magnitude at large pressures. Since the ultrathin
lubricant film consists of less than three molecular layers the relaxation
process of the temperature to the value $T_e$ occurs during time
satisfying condition $\tau_T\ll \tau_{\sigma}$. Then, within the adiabatic approximation
$\tau_{\sigma} \gg \tau_{\varepsilon},~ \tau_{T}$, equations (\ref{eq3}) and (\ref{eq4}) are reduced to the
time dependencies
\begin{eqnarray} \varepsilon(t)&=&\bar \varepsilon + \tilde\varepsilon\xi_4(t),\quad
T(t)=\bar T + \widetilde T\xi_5(t); \label{X1}\\
\bar \varepsilon &{\equiv}& \sigma^a \left(T_{e} {-} 1 {+} \sigma^{2a}\right) d_a(\sigma),~
\tilde{\varepsilon}{\equiv}\sqrt{I_\varepsilon {+} I_T \sigma^{2a}}
~d_a(\sigma), \nonumber \\
\bar T &{\equiv}&\left(T_{e} {+} 2\sigma^{2a}\right) d_a(\sigma), ~
\widetilde{T}\equiv\sqrt{I_T {+} I_\varepsilon\sigma^{2a}}~d_a(\sigma), ~
d_a(\sigma)\equiv(1+\sigma^{2a})^{-1}. \label{X2}
\end{eqnarray}
Here, deterministic components are reduced to expressions obtained in Ref.~\cite{liq},
whereas fluctuational ones follow from the property known as
variance additivity of independent Gaussian random quantities
\cite{Risken}. Thus, the use of the slaving principle inherent in
synergetics \cite{13} transforms initially adiabatic noises both of strain
$\varepsilon$ and temperature $T$ to multiplicative form.  As a result, a combination of
Eqs.~(\ref{eq2}), (\ref{X1}), and (\ref{X2}) leads to the Langevin
equation \cite{stoh,dissipative,pla}:
\begin{equation}
\dot\sigma = f_a(\sigma)+\sqrt{I_a(\sigma)}~\xi(t),
\label{langevin}
\end{equation}
where the time $t$ is measured in the units of stress relaxation time $\tau_\sigma$.
Generalized force  $f_a(\sigma)$ and effective intensity of noise $I_a(\sigma)$ are fixed by equations \cite{stoh,dissipative,pla}:
\begin{eqnarray}
f_a(\sigma) &\equiv& - \sigma^a + g\sigma^a\left[1 - (2 - T_e)(1 + \sigma^{2a})^{-1}\right], \nonumber \\
I_a(\sigma) &\equiv& I_\sigma + g^2(I_\varepsilon + I_T\sigma^{2a})(1 + \sigma^{2a})^{-2}. \label{7}
\end{eqnarray}
Effective intensity of noise is obtained in accordance with variance additivity property of noise mentioned above.
In order to avoid mistakes, one should notice that a direct insertion of Eqs.~(\ref{X1}) and (\ref{X2}) into (\ref{eq2}) results in the appearance of a stochastic addition
\begin{equation}
\left[I_\sigma^{1/2} + \left(I_\varepsilon^{1/2} + I_T^{1/2}\sigma^a\right)gd_a(\sigma)\right]\xi(t),
\label{X4}
\end{equation}
whose squared amplitude is quite different from the effective noise intensity (\ref{7}).
Moreover, in contrast to the expressions (\ref{X2}), a direct use of the
adiabatic approximation in Eqs.~(\ref{eq3}) and (\ref{eq4}) reduces the fluctuational
additions in Eqs.~(\ref{X1}) to the forms:
$\tilde{\varepsilon}\equiv(I_\varepsilon^{1/2}+I_T^{1/2}\sigma^a)
d_a(\sigma)$, $\widetilde T\equiv(I_T^{1/2} - I_\varepsilon^{1/2}\sigma^a)d_a(\sigma)$.
The latter is obviously erroneous since the
effective noise of the temperature $\widetilde T$ disappears entirely
for the stress $\sigma=(I_T/I_\varepsilon)^{1/2a}$.
The reason for such a contradiction is caused by the fact that Langevin equation
does not permit the use of usual analysis methods (see \cite{Risken}).


Langevin equation (\ref{langevin}) is a stochastic differential equation (SDE), since it contains stochastic
force $\sqrt{I_a(\sigma)}~\xi(t)$. Therefore each solution of the equation is individual and we can say only about statistical characteristics of such solutions.
In this context further we consider only probability distribution of solutions $P_a(\sigma)$ over stress value $\sigma$.

Multiplying (\ref{langevin}) by $dt$, the Langevin differential relationship is obtained:
\begin{equation}
d\sigma = f_a(\sigma)dt + \sqrt{I_a(\sigma)}dW(t),
\label{langevin_diff}
\end{equation}
where $dW(t)=W(t+dt) - W(t)\equiv\xi(t)dt$ is the Wiener process with properties \cite{gard}:
\begin{equation}
\langle dW(t)\rangle = 0;\quad \langle(dW(t))^2\rangle = 2dt.
\label{winner_moment}
\end{equation}
In general case, infinite number of the FPE`s forms can correspond to equation (\ref{langevin_diff}).

There are several forms of interpretation, each can be characteristic of specific physical object. Three the most used are the Ito
interpretation (I-form), the Stratonovich interpretation (S-form) \cite{13} and the kinetic form (K-form) \cite{Klimontovich}. Within the framework of the Ito form stochastic processes $\sigma(t)$ and $dW(t)$ presented in the last term of equation (\ref{langevin_diff}) are supposed to be statistically independent \cite{13}. Integrating (\ref{langevin_diff}) with the use of the Stratonovich interpretation it is necessary to evaluate the last term in the center of the time interval, i.e., to use the following construction \cite{13,gard}:
\begin{equation}
\sqrt{I_a\left(\sigma\left(\frac{t_{i}+t_{i-1}}{2}\right)\right)}dW(t_i).
\label{IntStr}
\end{equation}
The appearance of correlation between processes $\sigma$ and $dW$ can be seen in this case suggesting about the presence of memory
effects. Such effects are often present in real systems. In other words the Stratonovich form corresponds to the equation (\ref{langevin_diff})
with real noise, that can be approximated by the Gaussian white noise. In general case the Ito form is used for biological systems with discrete
time \cite{Horstemke}, for statistical interpretation of birth-death processes of living organisms for example. Thus, for description of
melting of ultrathin lubricant film, one would rather choose the Stratonovich form, since in this case the time is continuous and correlations are present in the noise.

\begin{figure}[htbp]
\centering{\resizebox{8cm}{!}{\includegraphics{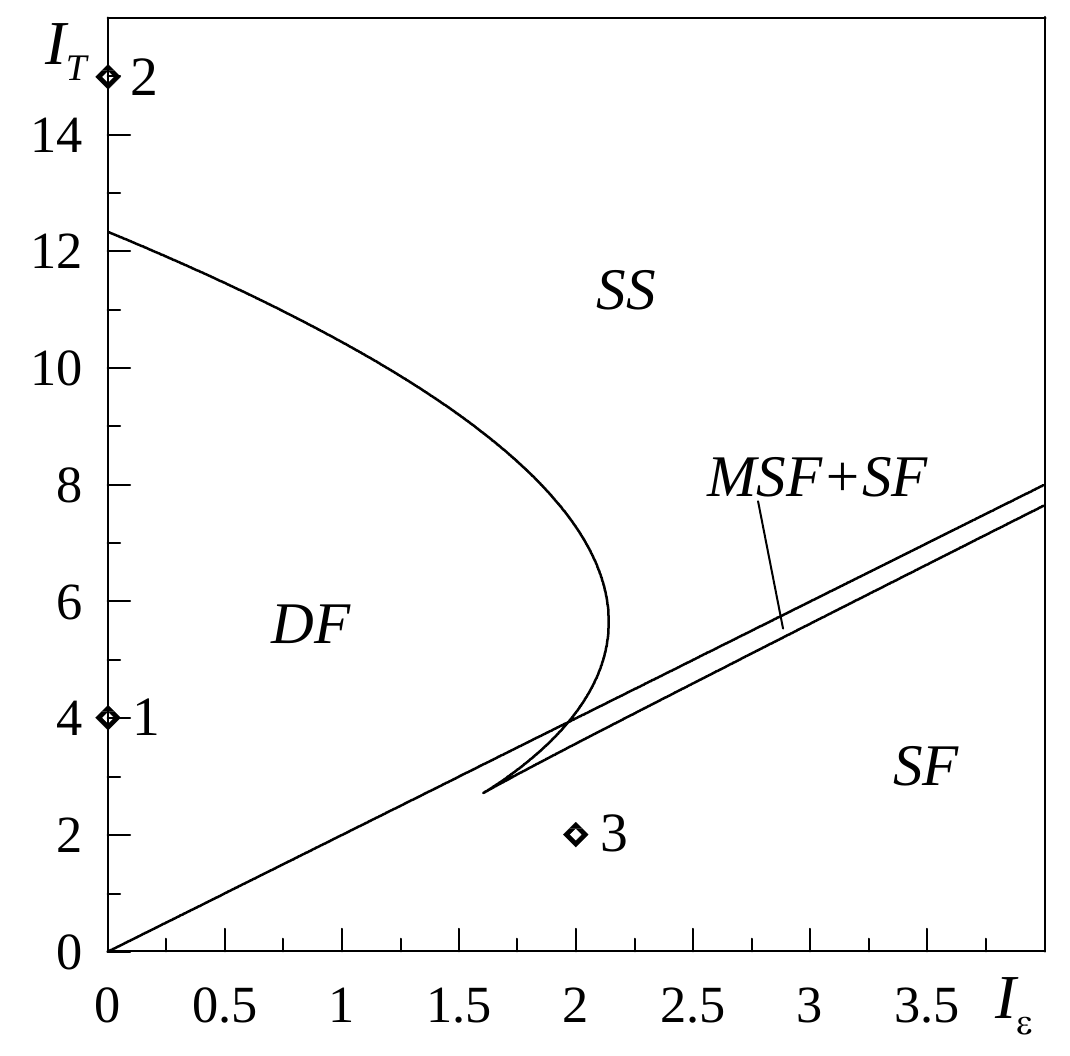}}}
\caption{Phase diagram at $g = 0.9, T_e = 1.5, a = 0.75$ with domains of  friction such as sliding ($\it SF$), dry ($\it DF$),
$stick-slip$ ($\it SS$), metastable and stable sliding ($\it MSF+SF$).}
\label{fig1}
\end{figure}

In the works \cite{stoh,dissipative} the simple Ito form of the FPE was used. Here within the framework of the Stratonovich form, we show that for our system the change of the form of interpretation does not lead to qualitative changes of its behavior.
The corresponding FPE with respect to (\ref{winner_moment}) is (S-form):
\begin{equation} \label{dFok_Plank_Strat}
\frac{\partial P_a(\sigma,t)}{\partial t}=
-\frac{\partial}{\partial\sigma}\left[f_a(\sigma) P_a(\sigma,t)\right] + \frac{\partial}{\partial\sigma}\left[
\sqrt{I_a(\sigma)}\frac{\partial}{\partial\sigma}\sqrt{I_a(\sigma)} P_a(\sigma,t)\right].
\end{equation}
Distribution of the solutions of (\ref{langevin_diff}) becomes stationary in time, and its form can be found from
(\ref{dFok_Plank_Strat}) at $\partial P_a(\sigma,t)/\partial t = 0$:
\begin{equation}
P_a(\sigma)={\mathcal Z}^{-1}\exp\lbrace-U_a(\sigma)\rbrace.
\label{a}
\end{equation}
The obtained distribution is defined by a normalization constant $\mathcal Z$ and an effective potential
\begin{equation}
U_a(\sigma)=\frac{1}{2}\ln I_a(\sigma)-\int\limits^\sigma_{0}\frac{f_a(\sigma')}{I_a(\sigma')}{d}\sigma'.
\label{ef_potential}
\end{equation}
Extremum points of the distribution (\ref{a}) are defined by condition $dU_a/d\sigma\equiv dI_a/d\sigma-2f_a=0$, or in explicit form
\begin{equation}
\frac{T_e - 2}{1 + \sigma^{2a}} + \frac{ag\sigma^{a-1}}{\left( {1 + \sigma ^{2a}} \right)^3}\left[2I_\varepsilon - I_T \left(
{1 - \sigma ^{2a}} \right) \right] = \frac{1 - g}{g}. \label{maxx}
\end{equation}
So, extremum abscissas of $P_a(\sigma)$ are independent of noise intensity $I_\sigma$. Expression (\ref{maxx})
differs from analogous one, obtained in \cite{stoh}. In \cite{stoh} the second term is multiplied by $2a$, but in this case it
is multiplied by $a$. Thus, at increase in all noises intensities by two times further examination\footnote{We study the extremums of distributions for phase diagrams analysis and interpretation of the stationary states.} within the framework of the Stratonovich interpretation concurs with results, obtained using the Ito form \cite{stoh}. However,
potential (\ref{ef_potential}) does not take earlier obtained form \cite{stoh} at simple renormalization of the noise intensities,
since it differs from above only by the first term (presence of factor 1/2). Therefore the time dependencies of the stresses are different.
Since the aim of this work is to study peculiarities of time evolution of the stress, we use the Stratonovich approach. Earlier studies
\cite{stoh} were focused only on stationary states using the Ito approach. The typical phase diagram at fixed temperature $T_e$ is shown in fig.~\ref{fig1}, where lines correspond to the stability loss limits of the system.
Straight line going from the beginning of coordinates is defined by
\begin{equation}
I_T=2I_\varepsilon.
\label{granitsa}
\end{equation}
This follows from (\ref{maxx}) and limits existence of zero stationary solution $\sigma_0=0$. Above this line the maximum of
$P_a(\sigma)$ always exists at $\sigma_0=0$, below it the maximum does not exist. In the diagram four domains with different regimes of friction can be seen.

\begin{figure}[htbp]
\centering{\resizebox{8cm}{!}{\includegraphics{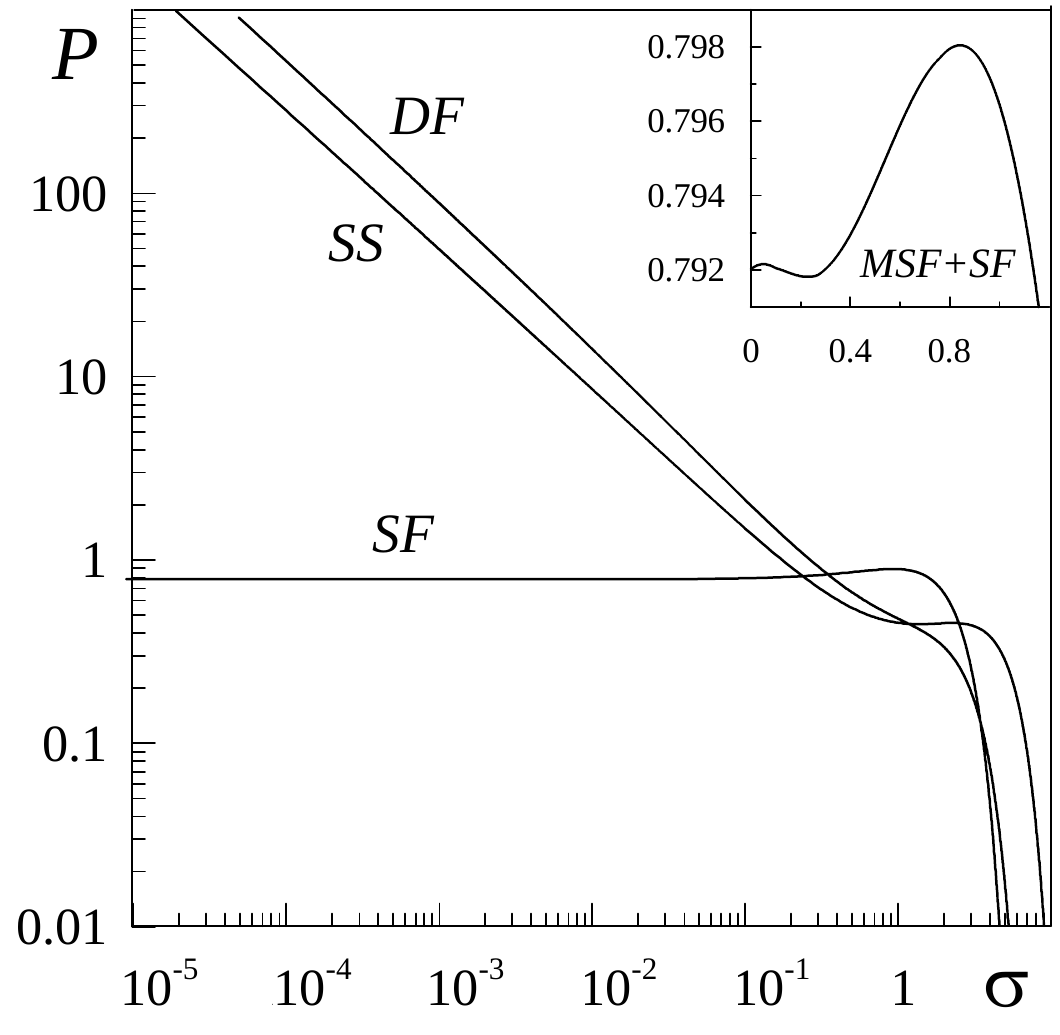}}}
\caption{Distribution (\ref{a}) at $I_\sigma=10^{-10}$ and regimes shown by points in fig.~\ref{fig1}:
1 --- $I_\varepsilon = 0, I_T = 4$ ($\it DF$);
2 --- $I_\varepsilon = 0, I_T = 15$ ($\it SS$);
3 --- $I_\varepsilon = 2, I_T = 2$ ($\it SF$).
The insert shows $P_a(\sigma)$ at $I_\varepsilon = 1.968, I_T = 3.5$ ($\it MSF+SF$).}
\label{fig2}
\end{figure}

Unnormalized probability distribution (\ref{a}) shown in fig.~\ref{fig2} corresponds to domains in fig.~\ref{fig1}.
Point 1 is located in dry friction region ($\it DF$) of the phase diagram, and single maximum of distribution is observed at $\sigma_0=0$. Two-phase region $\it SS$ is defined by existence of distribution maximums $P_a(\sigma)$ at zero and non-zero stress values (point 2).
Point 3 is located in domain, where only one distribution maximum exists at $\sigma_0\ne 0$, being related to liquid friction or
sliding regime ($SF$). In the insert distribution in area $\it MSF+SF$ is shown. Dependence $P_a(\sigma)$ has two maximums at $\sigma_0\ne 0$ corresponding to interrupted regime when transitions between stable and metastable sliding friction are possible.

\begin{figure}[htbp]
\centering{\resizebox{8cm}{!}{\includegraphics{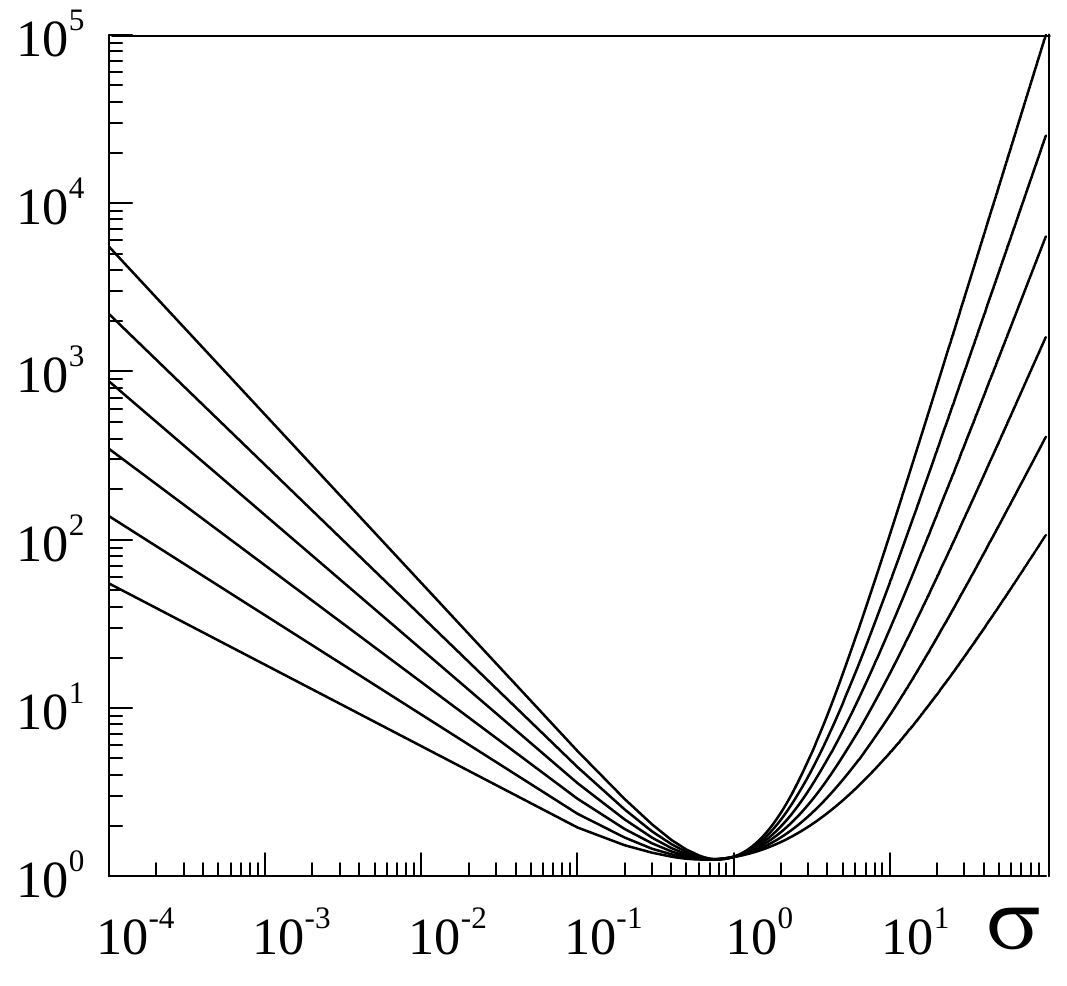}}}
\caption{Integrand of (\ref{15}) at the same parameters as in fig.~\ref{fig1} and $a = 0.5, 0.6, 0.7, 0.8, 0.9, 1.0$ from bottom to top.}
\label{fig3}
\end{figure}

The $P_a(\sigma)$ dependencies in fig.~\ref{fig2} are plotted in log-log coordinates. It is  seen that for curves $\it DF$ and $\it SS$
distribution takes the power-law form. Such regime corresponds to values $\sigma \ll 1$ and $I_\sigma, I_\varepsilon \ll I_T$,
at which (\ref{a}) is written as follows
\begin{equation}
P_a(\sigma) = \sigma^{-a}{\mathcal P}(\sigma),
\label{14}
\end{equation}
where ${\mathcal P}(\sigma)$ is defined by:
\begin{eqnarray}
{\mathcal P}\left( \sigma \right) &=& {\mathcal Z}^{-1}g^{-1}I_T^{-1/2}(1+\sigma^{2a})\times\nonumber\\
&\times&\exp\left\{-I_T^{-1}g^{-2}\int\limits_0^\sigma \frac{1 - g\left[1 - (2-T_e)(1+(\sigma')^{2a})^{-1}\right]}
{(1+(\sigma')^{2a})^{-2}(\sigma')^a}d{\sigma}'\right\}.
\label{15}
\end{eqnarray}
It is known that self-similar systems have a homogenous distribution \cite{Amit}. Distribution (\ref{14}) becomes homogenous at constant
function (\ref{15}). At small stress values the multiplier before exp is $1+\sigma^{2a}\to 1$. In fig.~\ref{fig3}
the integration element is plotted without coefficient before integral.
As can be seen from the figure integral in (\ref{15}) has small value at $\sigma<0.8$, and when $\sigma$ exceeds certain value it rapidly
begins to increase. According to structure of the equations (\ref{14}), (\ref{15}), integral gives the basic contribution in the resultant
distribution (\ref{14}), which becomes exponentially decreasing. Value $\sigma\approx 0.8$ conforms with fig.~\ref{fig2}. Thus, power-law distribution, typical for self-similar behavior, exists in limited range of stress values. Self-similar properties disappear when the stress exceeds the critical value.

Stratonovich approach leads to first important difference from Ito form: in (\ref{14}) distribution exponent is equal to $-a$,
while in \cite{stoh} it is $-2a$.

\section{Stress time series}

\begin{figure}[htbp]
\hspace{2cm}
\resizebox{12cm}{!}{\includegraphics{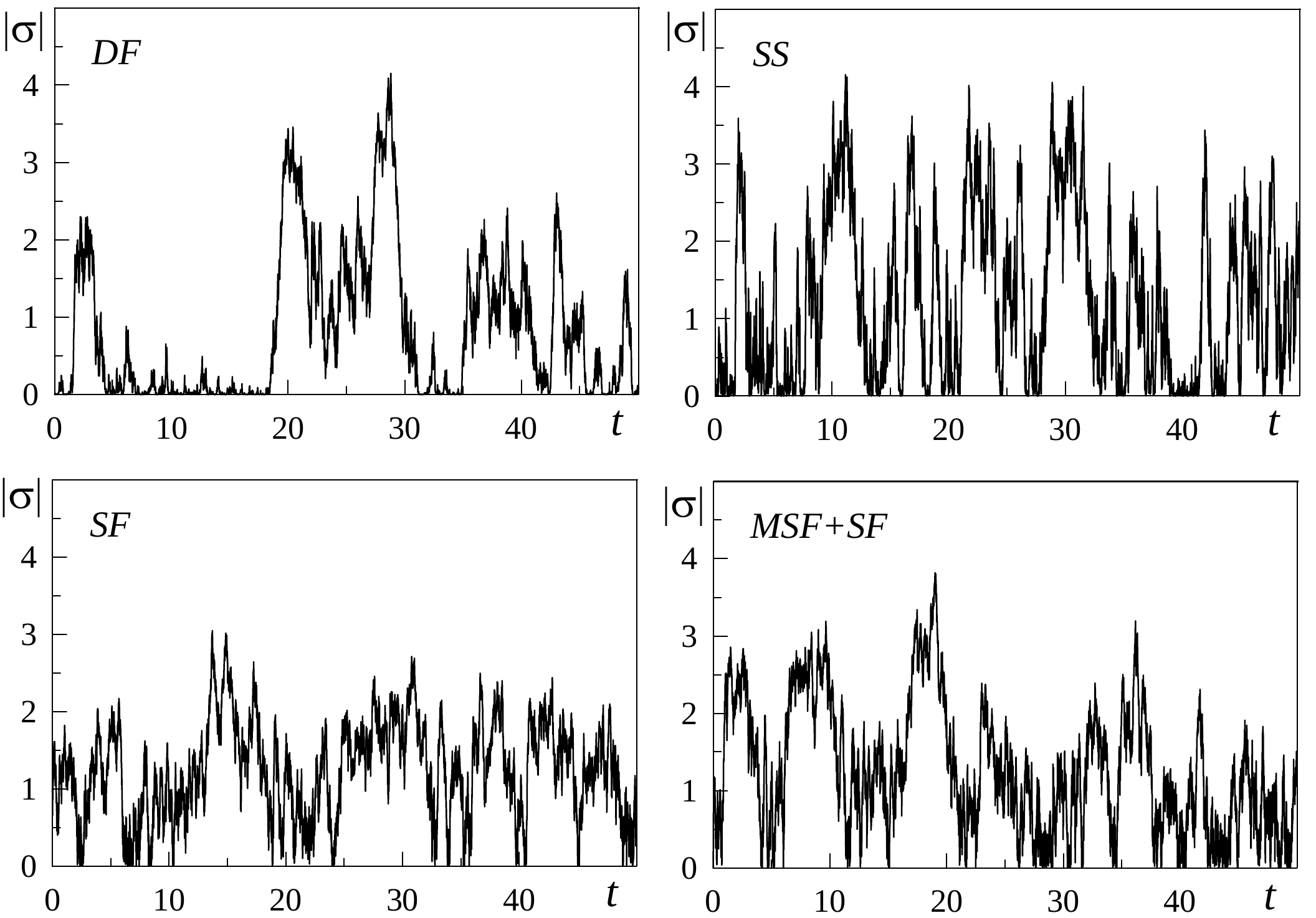}}
\vspace{1cm}
\caption{Stress time series  $|\sigma|(t)$, derived from equation (\ref{ito}) by numerical solution according to (\ref{iter})
at $N=10^4, t=50, dt=0.005$. Regimes, that are shown in the plot, correspond to points in phase diagram (fig.~\ref{fig1}).}
\label{fig4}
\end{figure}

\begin{figure}[!ht]
\centerline{\includegraphics[width=140mm]{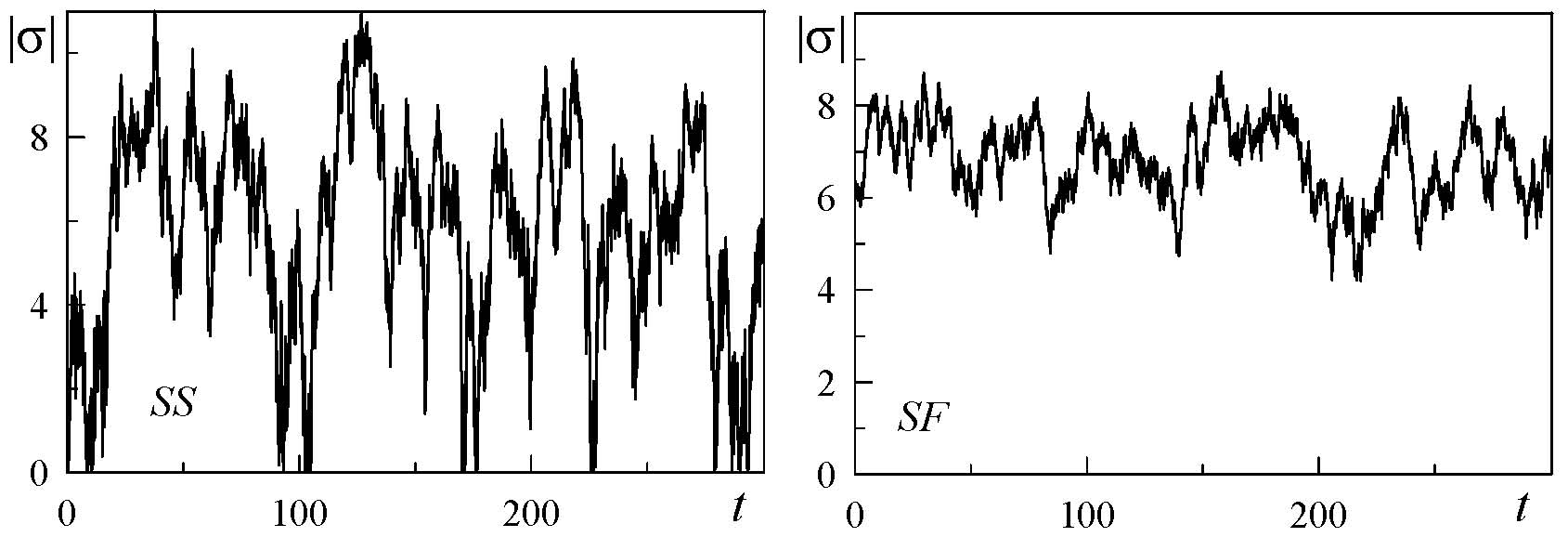}}
\caption{Stress time series $|\sigma|(t)$ corresponding to fig.~\ref{fig4} at $T_e=4$.}
\label{fig5}
\end{figure}

\begin{figure}[htbp]
\centering{\resizebox{8cm}{!}{\includegraphics{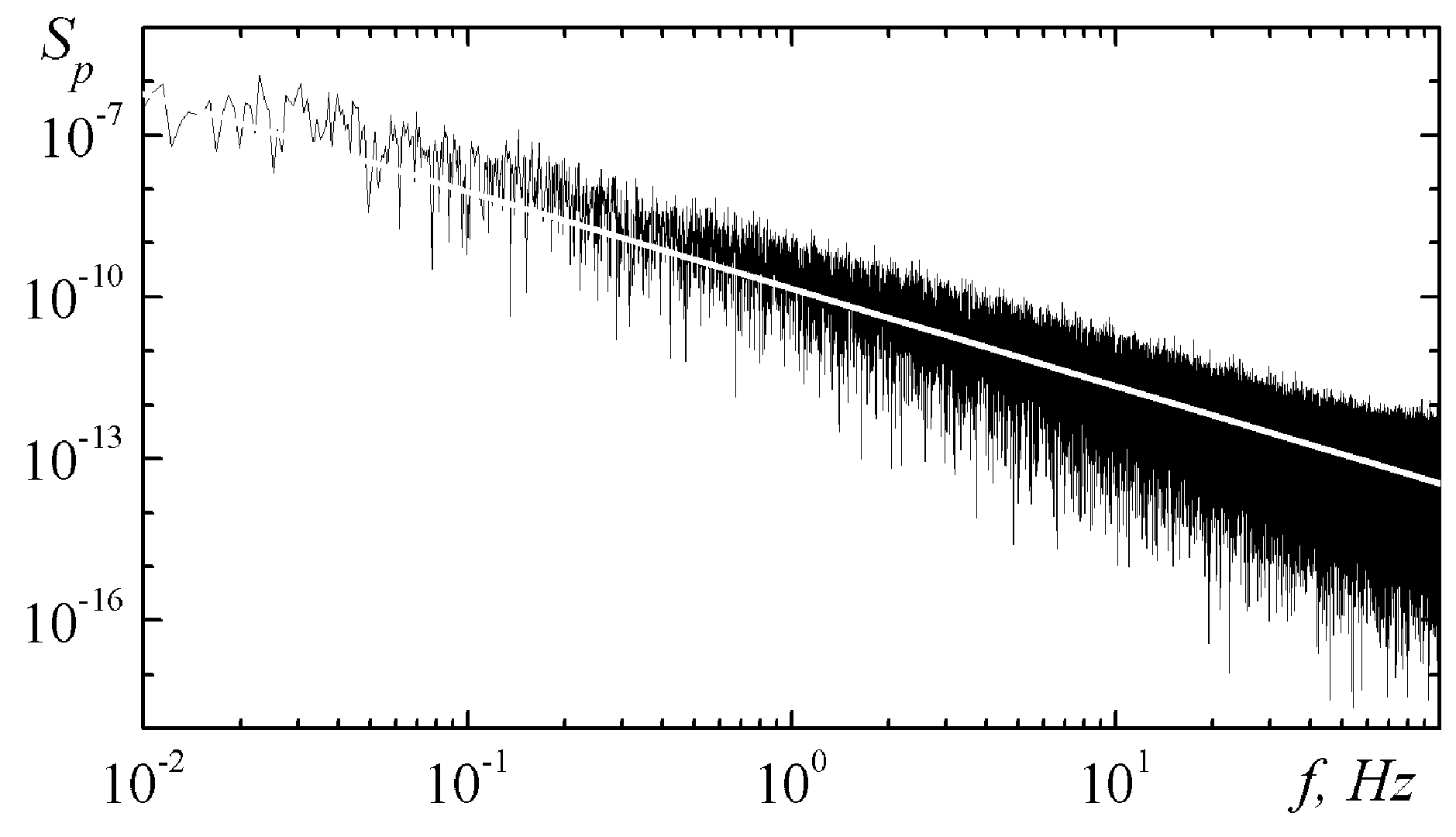}}}
\caption{Oscillation spectrum $S_p(f)$ corresponding to the $\it SS$ regime parameters shown in fig.~\ref{fig4}. White line is fixed by equation $S_p(f)\propto 1/f^{1.8}$. Power $S_p$ is measured in conventional units.}
\label{fig6}
\end{figure}

Euler method is used for numerical solution of the equation (\ref{langevin_diff}). Iterative procedure differs from analogous one used in \cite{dissipative} because Eq.~(\ref{langevin_diff}) is the Stratonovich`s SDE. To use common iterative procedure it is necessary to transform Stratonovich`s SDE to the equivalent Ito`s SDE. Taking into account the properties~(\ref{winner_moment}) for Eq.~(\ref{langevin_diff}) one obtains the following form of the Ito`s SDE \cite{gard,Horstemke}:
\begin{equation}
d\sigma = \left[f_a(\sigma)+
\sqrt{I_a(\sigma)}\frac{\partial}{\partial\sigma}\sqrt{I_a(\sigma)}\right]dt + \sqrt{I_a(\sigma)}dW(t).
\label{ito}
\end{equation}
According to definition of the discrete analog of stochastic force differential $dW(t) \equiv \sqrt{\Delta t}W_i$ and (\ref{7}),
one can obtain iterative procedure for solution of the equation (\ref{ito}):
\begin{equation}
\sigma_{i+1} = \sigma_i + \left(f_a(\sigma_i)+\frac{ag^2\sigma_i^{2a-1}[I_T(1-\sigma_i^{2a})-2I_\varepsilon]}
{(1+\sigma_i^{2a})^3}\right)\Delta t + \sqrt{I_a(\sigma_i)\Delta t}W_i.
\label{iter}
\end{equation}
Solution of the equation runs over $t\in[0,T]$ time interval. At given numbers of iterations $N$ (number of time series members) increment
of time is defined as $\Delta t = T/N$. Force $W_i$ has
following peculiarities (cf.~(\ref{winner_moment})):
\begin{equation}
\langle W_i \rangle=0, \quad \langle W_i W_{i'} \rangle = 0, \quad
\langle W^2_i \rangle \to 2.
\label{Sila_diskret}
\end{equation}
The Box-Muller model allows us to represent sufficiently stochastic force \cite{C++}:
\begin{equation}
W_i = \sqrt{\mu^2}\sqrt{-2\ln r_1}\cos(2\pi r_2), \quad r_n \in (0,1],
\end{equation}
where, according to (\ref{Sila_diskret}), $\mu^2=2$ is the dispersion,   $W_i$ is the random number with properties (\ref{Sila_diskret}).
Pseudo-random numbers $r_1$, $r_2$ have uniform distribution and repeat themselves through periodical intervals. Effective potential
(\ref{ef_potential}) has minimums at positive and negative values of stress $\sigma$. Thus while solving numerically Eq.~(\ref{ito}) fluctuations cause transitions between states defined by the  minimums. We can exclude negative part $\sigma<0$ out of consideration since one-directional motion of upper moving
surface is considered. This allows us to analyze further the behavior of $|\sigma|(t)$. Typical realizations of the $|\sigma|(t)$ for considered regimes are shown in fig.~\ref{fig4}.
Positive domains whith stress value close to zero is observed at dry friction regime ($\it DF$). There are random transitions between zero and non-zero stress values $\sigma$ at $stick-slip$ regime ($\it SS$). Realizations of the $\it SF$ and the $\it MSF+SF$ regimes are visually
similar. Therefore to detect the friction type one needs to apply additional analysis for probability density definition
(see~fig.~\ref{fig2}). Time series obtained in work \cite{dissipative} permit visual interpretation because corresponding phase diagrams were plotted in $T_e-I_T$ coordinates, and $\sigma(t)$ were built at different values of friction surfaces temperature $T_e$. At large temperatures $T_e$ lubricant is totally melted, at small $T_e$ it is solid.
Here, phase diagram is plotted at fixed value $T_e$, therefore the time series related to different regimes are similar. They
represent different friction regimes according to probability distributions shown in fig.~\ref{fig2}. Note, plotting the phase
diagram shown in fig.~\ref{fig1}, it is unreasonable to use large $T_e$ values because it is transformed into straight line (\ref{granitsa}) demarcating the $\it SS$ and the $\it SF$ friction regimes, and other domains are eliminated.
For comparison, realizations of  $|\sigma|(t)$ are shown
in fig.~\ref{fig5} at the same parameters as in fig.~\ref{fig4}, but at $T_e=4$.
It can be defined visually that shown dependencies are in accordance with the $\it SS$ and the $\it SF$ regimes.
There are transitions between zero and non-zero stresses values in the $\it SS$ regime and in the $\it SF$ regime always $\sigma>0$.
However, the aim of this work is the analysis of the self-similar behavior, and we study all possible friction regimes. Therefore we use dependencies
shown in fig.~\ref{fig4}. In fig.~\ref{fig6} spectrum of the stress oscillations is shown, obtained by the fast Fourier transform algorithm
(FFT) \cite{C++} at the $\it SS$ regime time series analysis presented in fig.~\ref{fig4}.
Corresponding time series are obtained by iterative procedure (\ref{iter}) at $N=2{\cdot}10^5, t=10^3, dt=0.005$. It is evident that
signal power in the spectrum is decreased with increase in frequency. White line described by relationship $S_p(f)\propto 1/f^{1.8}$
is the spectrum approximation, i.e., power is inversely proportional  to frequency. Thus there are different time correlations in the system,
in contrast to the white noise for which  $S_p(f)=\rm const$. For all considered regimes spectrums $S_p(f)$ have similar form,
and for all cases $S_p(f)\propto 1/f^{1.8}$. Thus system considered on the basis of equation (\ref{langevin}) transforms white noise generators inherent in almost in all physical models into color noise with non-zero correlation time. Such behavior was observed experimentally \cite{skokov}.

\section{Multifractal detrended fluctuation analysis (MF-DFA)}

Multifractal analysis allows to calculate numerically the basic multifractal characteristics \cite{frac} describing the self-similar
systems. This method was proposed and developed by J. Kantelhardt et al. \cite{cantel}, and it is widely used in many scientific fields
for different time series analysis (meteorology \cite{meteo}, medicine \cite{medic}, economy\cite{econom},
and others \cite{struct,sunspot,dros}).
Here we cite from original work \cite{cantel} main statements of the method (for full description see \cite{cantel}): 
Supposing that $x_k$ is a series of length $N$, to provide the multifractal detrended fluctuation analysis, one must
follow next five steps:
\begin{itemize}
\item \textit{ Step 1}: Determine the "profile"
\begin{equation}
\label{eqp1} Y(i) = \sum\limits_{k = 1}^i {\left[ {x_k -  \langle x \rangle}
\right]} , \qquad \qquad i=1,...,N
\end{equation}
\item \textit{ Step 2}: Divide the profile $Y(i)$ into $N_s \equiv int(N/s)$ non-overlapping segments of equal lengths $s$.
Since the length $N$ of the series is often not a multiple of the considered time scale $s$, a short part at the end of the
profile may remain. In order not to disregard this part of the series, the same procedure is repeated starting from the opposite
end. Thereby, $2{N}_{s}$ segments are obtained altogether.
\item \textit{ Step 3}: Calculate the local trend for each of the $2{N}_{s}$ segments by a least-square fit of the series.
Then determine the fluctuation function
\begin{equation}
\label{eqp2} F^2(\nu ,s) = \frac{1}{s}\sum\limits_{i = 1}^s {\left\{
{Y\left[ {\left( {\nu - 1} \right)s + i} \right] - y_\nu (i)}
\right\}} ^2
\end{equation}
for each segments $\nu$, $\nu = 1,...,N_s$, and
\begin{equation}
\label{eqp3} F^2(\nu ,s) = \frac{1}{s}\sum\limits_{i = 1}^s {\left\{
{Y\left[ {N - \left( {\nu - N_s } \right)s + i} \right] - y_\nu (i)}
\right\}} ^2
\end{equation}
for $\nu = N_s + 1,...,2N_s $. Here, $y_{\nu}(i)$ is the fitting polynomial in segment $\nu$. Order of polynomial $m$ selected
with respect to the order of trend presenting in the series. Thus, polynomial of the $m$ order can eliminate trend of order $m-1$.
\item \textit{ Step 4}: Average over all segments to obtain the $q-th$ order fluctuation function:
\begin{equation}
\label{eqp4} F_q (s) = \left\{ {\frac{1}{2N_s }\sum\limits_{\nu =
1}^{2N_s } {\left[ {F^2(\nu ,s)} \right]^{q / 2}} } \right\}^{1/q},
\end{equation}
where, the index $q$ can take any real value except zero.
\item \textit{ Step 5}: Determine the scaling behavior of the fluctuation function by analyzing the log-log plot
of $F_q(s)$ versus $s$ for each value of $q$. If the series $x_i$ are long-range power-law correlated, $F_q(s)$
will depend on $s$, as a power law,
\begin{equation}
F_q\sim s^{h(q)},
\label{scal}
\end{equation}
\end{itemize}
where $h(q)$ is the generalized Hurst exponent depending on $q$ (note that $h(q)$ at $q=2$ is equal to classic Hurst 
exponent $H$ \cite{feder}). 

\noindent Function $h(q)$ is connected with another classic multifractal
scaling exponent $\tau(q)$ \cite{cantel,frac}:
\begin{equation}
\tau(q)=qh(q)-1.
\label{tay}
\end{equation}
Self-similar behavior can be described by multifractal spectrum function $f(\alpha)$, connected with $\tau(q)$ through Legendre
transformation \cite{frac}:
\begin{equation}
\alpha=\tau'(q), \quad f(\alpha)=q\alpha-\tau(q),
\label{afa}
\end{equation}
where $\alpha$ is the Holder exponent, and "$'$" denotes differentiation with respect to $q$. Using (\ref{tay}), we can get directly
related $f(\alpha)$ and $h(q)$:
\begin{equation}
\alpha=h(q)+qh'(q), \quad f(\alpha)=q[\alpha-h(q)]+1.
\end{equation}
Type of denoted dependencies characterizes time series behavior. Thus, constant value of $h(q)=\rm const$ and, correspondingly, linear
increase in exponent $\tau(q)$ denote monofractal series. Decrease in $h(q)$ with $q$ and nonlinear growth of $\tau(q)$ are inherent in
multifractal time series. Just one value of the Holder exponent $\alpha$ is characteristic for monofractal objects,
and $f(\alpha)$ dependence presents a narrow peak. There is spectrum of $f(\alpha)$ values in the case of multifractal series. 
However, in the case of monofractal series the numerical calculation will not give the only value of $f(\alpha)$, instead 
we will have a set of close values $\alpha$, which, comparing with wider spectrums, approximately can be considered as monofractal issues.
In general case, two types of multifractality can be distinguished for time series: (1) multifractality caused by broad probability density
function of the series members, (2) multifractality caused by different time correlations between series members. Rearrangement of series components in a random order
does not lead to the elimination multifractality in the first case. In the second case randomization leads to disappearance of the correlations.
Since the reason for multifractality vanishes the series is transformed to monofractal. If both reasons of multifractality are inherent in series,
the corresponding mixed series is characterized by weaker multifractality than initial one \cite{cantel}. Thus, analyzing the mixed series using the method \cite{cantel}, it is obviously
possible to define the reason for multifractality and presence of time correlations.

In general case, for time series two types of multifractality can be distinguished \cite{cantel}: (1) multifractality caused by broad probability density
function of the members of the series, (2) multifractality caused by different range time correlations between series members. To define the reason for multifractality and presence of time correlations, one must apply shuffling procedure that consist in rearranging components of series in a casual order, and then compare corresponded spectrums $f(\alpha)$ of original and shuffled series. Thus, after shuffling of the series with multifractality of type (1) corresponded $f(\alpha)$ function (multifractality) must not changed, because the probability density remains the same. In the second case hashing leads to disappearance of available correlations, and since the reason for multifractality thus vanishes, such series is transformed to monofractal. If for a series both reasons of multifractality are inherent in, the corresponding mixed series are characterized by
weaker multifractality than initial series \cite{cantel}.

\begin{figure}[htbp]
\centering{\resizebox{9cm}{!}{\includegraphics{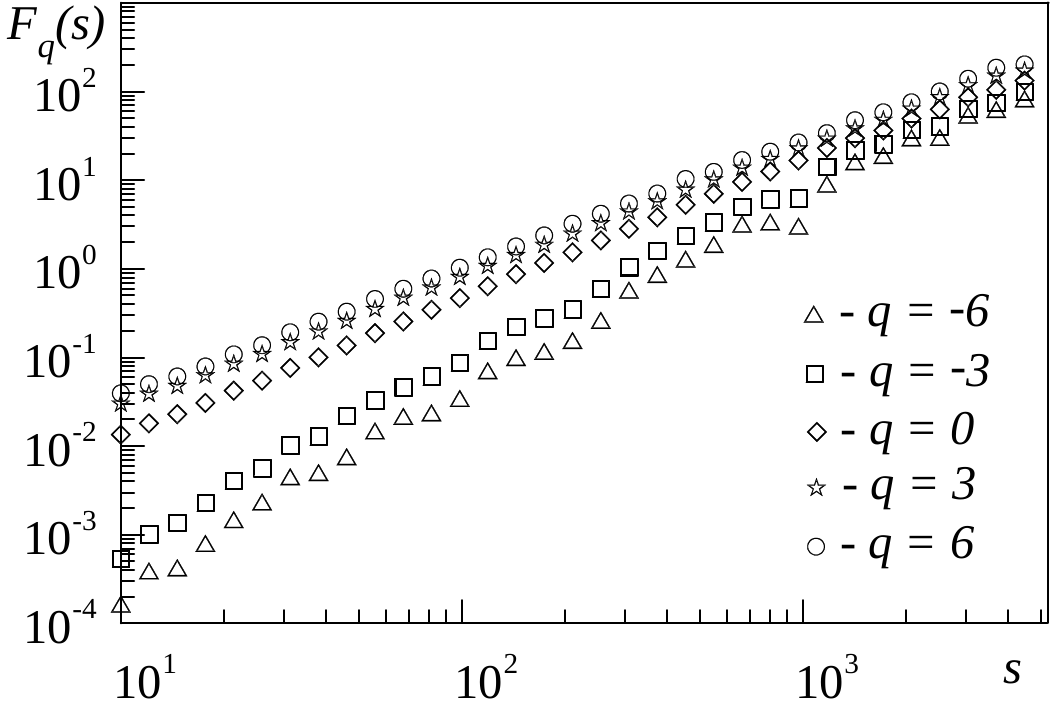}}}
\caption{Dependence of fluctuation function of $F_q(s)$ for the parameters of fig.~\ref{fig4} and the $\it DF$ regime.}
\label{fig7}
\end{figure}

Using this method we analyze stress time series $|\sigma|(t)$ shown in fig.~\ref{fig4}. Figure~\ref{fig7} illustrates typical
form of the $F_q(s)$ dependence, plotted in log-log coordinates, at some $q$ values for time series related to the $\it DF$ regime.
From this figure we can see linear dependence on all set of $s$ values, that is typical for all series analyzed in the current work.

This allows us to calculate precisely the Hurst exponent $h(q)$ according to the scaling equation (\ref{scal}). We select domain $50<s<500$
for calculation of the multifractal characteristics, where dependence of $F_q(s)$ has linear form.

\begin{figure}[htb]
\centerline{\includegraphics[width=130mm]{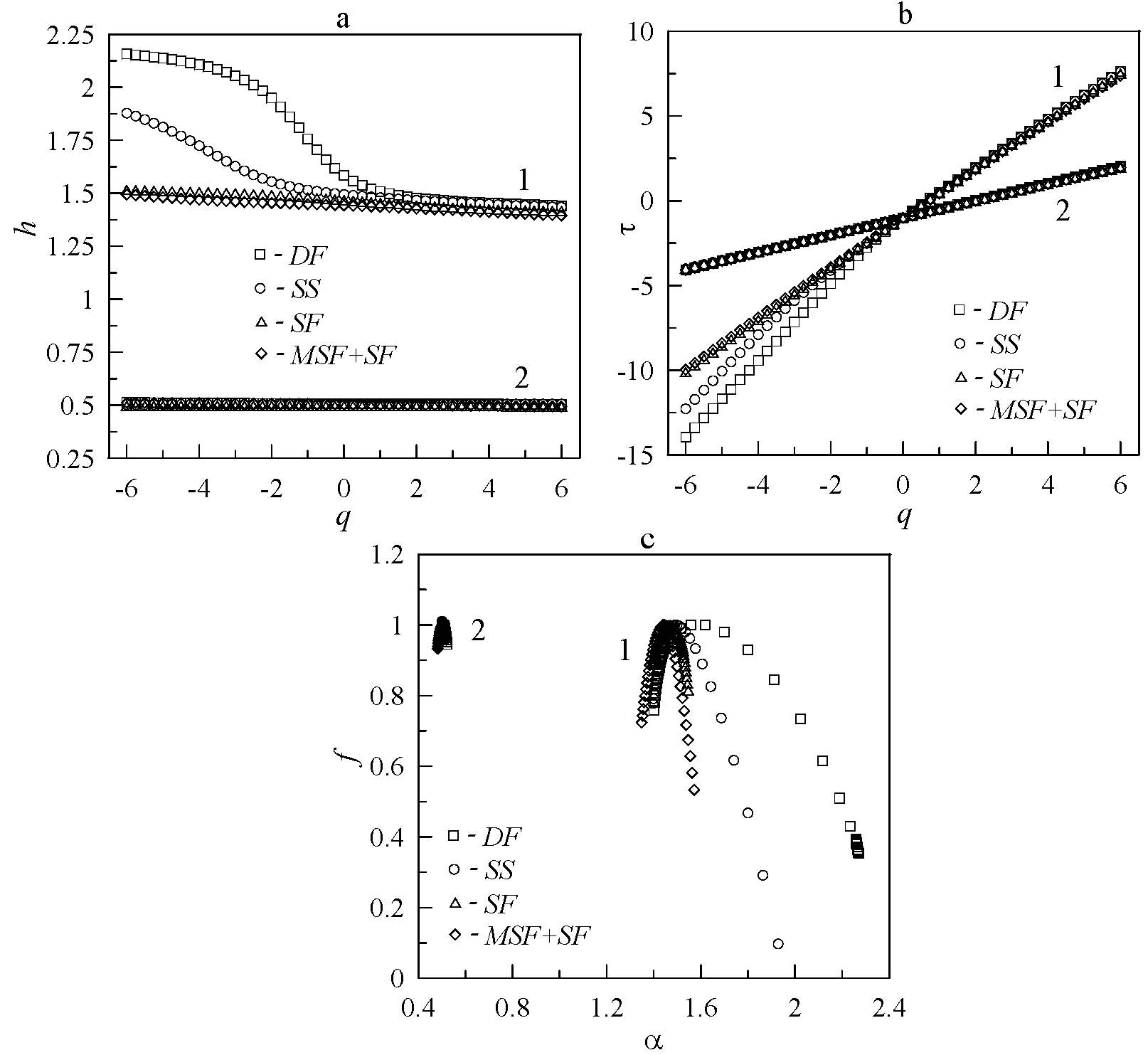}}
\caption{Multifractal characteristics $h(q)$, $\tau(q)$, and $f(\alpha)$ corresponding to the parameters of time series
in fig.~\ref{fig4}. Set of curves~1 is related to series, derived directly from procedure (\ref{iter}), and set~2 is related to
analogous shuffled series.}
\label{fig8}
\end{figure}

For time series shown in fig.~\ref{fig4} at $N=10^5, t=10^3, dt=0.01$ we can calculate $h(q)$, $\tau(q)$, and $f(\alpha)$. From
fig.~\ref{fig8} it is seen that the strongest multifractality is exhibited by series related to the $\it DF$ regime, then $\it SS$
follows, and for series related to the $\it MSF+SF$ and the $\it SF$ regimes the weaker dependence $h$ on $q$ is characteristic that
corresponds to monofractal behavior.
Strong multifractality for the $\it DF$ regime can be explained by power-law probability density function at small stresses, this
is inherent in self-similar systems. In the $\it SS$ regime multifractality is weaker, since probability density function of the Langevin equation solution also has non-zero maximum.
There is possibility of system transition into state defined by this maximum, related to lubricant
melting when it loses self-similar properties and settles into stationary regime of the sliding friction.
But, existence of two maximums of $P_a(\sigma)$ corresponds to the $stick-slip$ regime, and inverse transitions to
solid-like structure may occur, and system returns to self-similarity.

For the $\it MSF+SF$ and the $\it SF$ regimes system does not exhibit multifractality,
because of non-power-law probability density function.

\begin{figure}[!ht]
\centerline{\includegraphics[width=100mm]{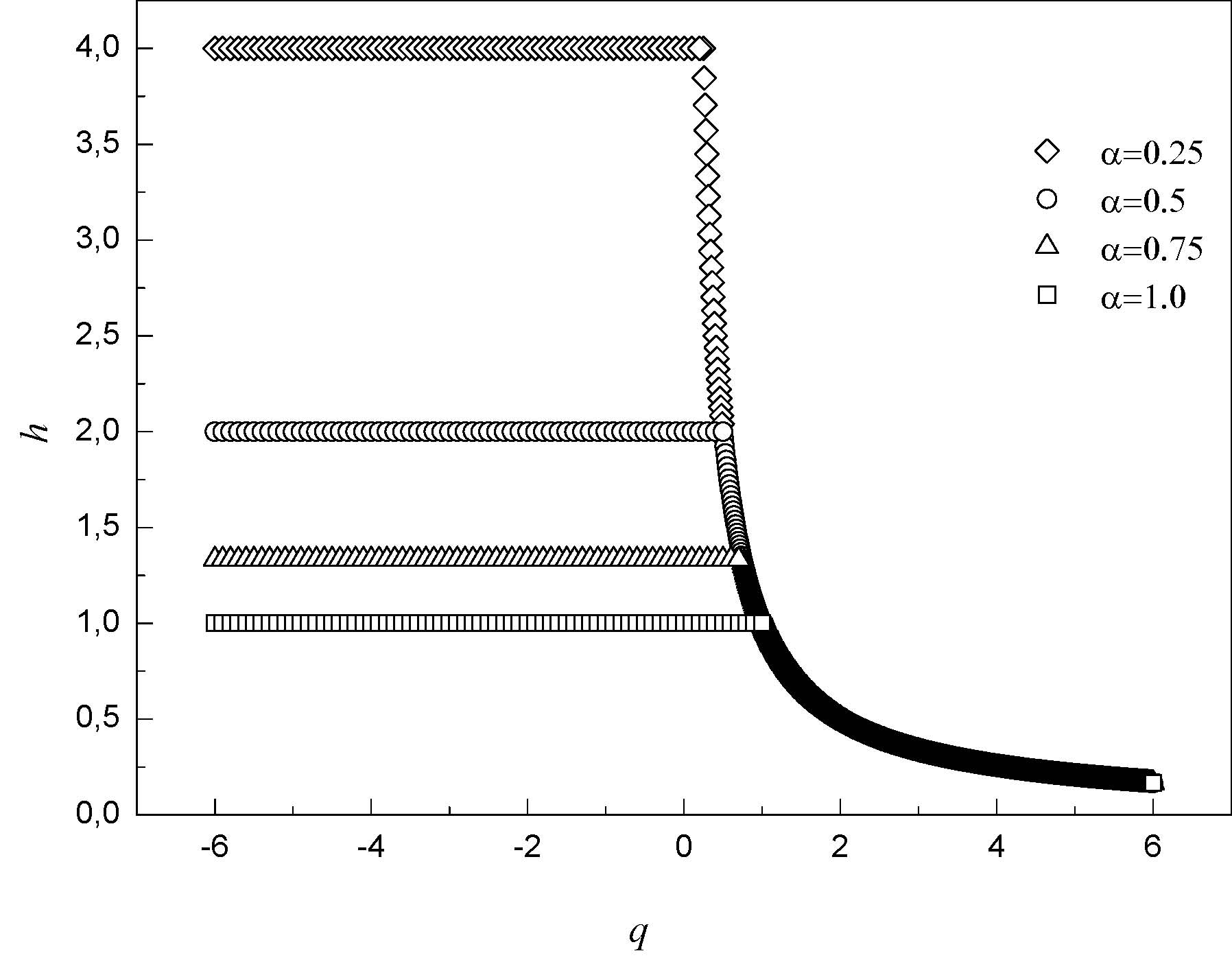}}
\caption{Hurst exponent $h(q)$ for series with power-law distribution (\ref{pwl}) obtained from eq.(\ref{mf}).}
\label{fig99}
\end{figure}

Peculiarity of the results shown in fig.~\ref{fig8} is that the multifractal characteristic for different friction regimes
has close values in the $q>0$ domain
and main difference observed in the area where $q$ less than zero. In that range of $q$ values MF-DFA takes into account small
fluctuations in the time series, so it may seems that curves shown in fig.~\ref{fig8} may differs due to the peculiarity of the
numerical realization of the MF-DFA procedure, and not due to the different statistical properties of the series, corresponded to
variant friction regimes as we mentioned above. To explain this situation we must note that such results are typical for series
with power-law distribution function. Thus, for uncorrelated multifractal
series with distribution function
\begin{equation}
P(x)=\alpha {x}^{-(\alpha+1)}, \label{pwl}
\end{equation}
where $\alpha>0$ and $1\leq x < \infty$, corresponded Hurst exponent $h(q)$ can be determined analytically \cite{cantel}, and defines as:
\begin{equation}
h(q)\sim\left\{
{\begin{array}{l}
1/q\quad (q>\alpha)\\
1/\alpha \quad(q\le\alpha).\\
\end{array}} \right.
\label{mf}
\end{equation}
This dependencies shown in fig.~\ref{fig99}.
Latter curves have the same topology as results for stress time series shown on fig.~\ref{fig8}, namely that main
difference between multifractal Hurst exponent corresponded to series with different power-law distributions observed
in $q<0$ domain. This is explicit result, characterful for time series with power-low distribution, and we suppose that
results of our calculations of multifractal characteristics for stress time series are precision and correct. We also
note that for stress time series magnitude of the corresponded $h(q)$ function not determined by power exponent like in
eq.(\ref{mf}), because in that case we have power-law distribution only for small stress values (for DF regime approximately
$\sigma < 1$), and for larger $\sigma$ it breaks (see fig.~\ref{fig2}).

\begin{figure}[htbp]
\centering{\resizebox{8cm}{!}{\includegraphics{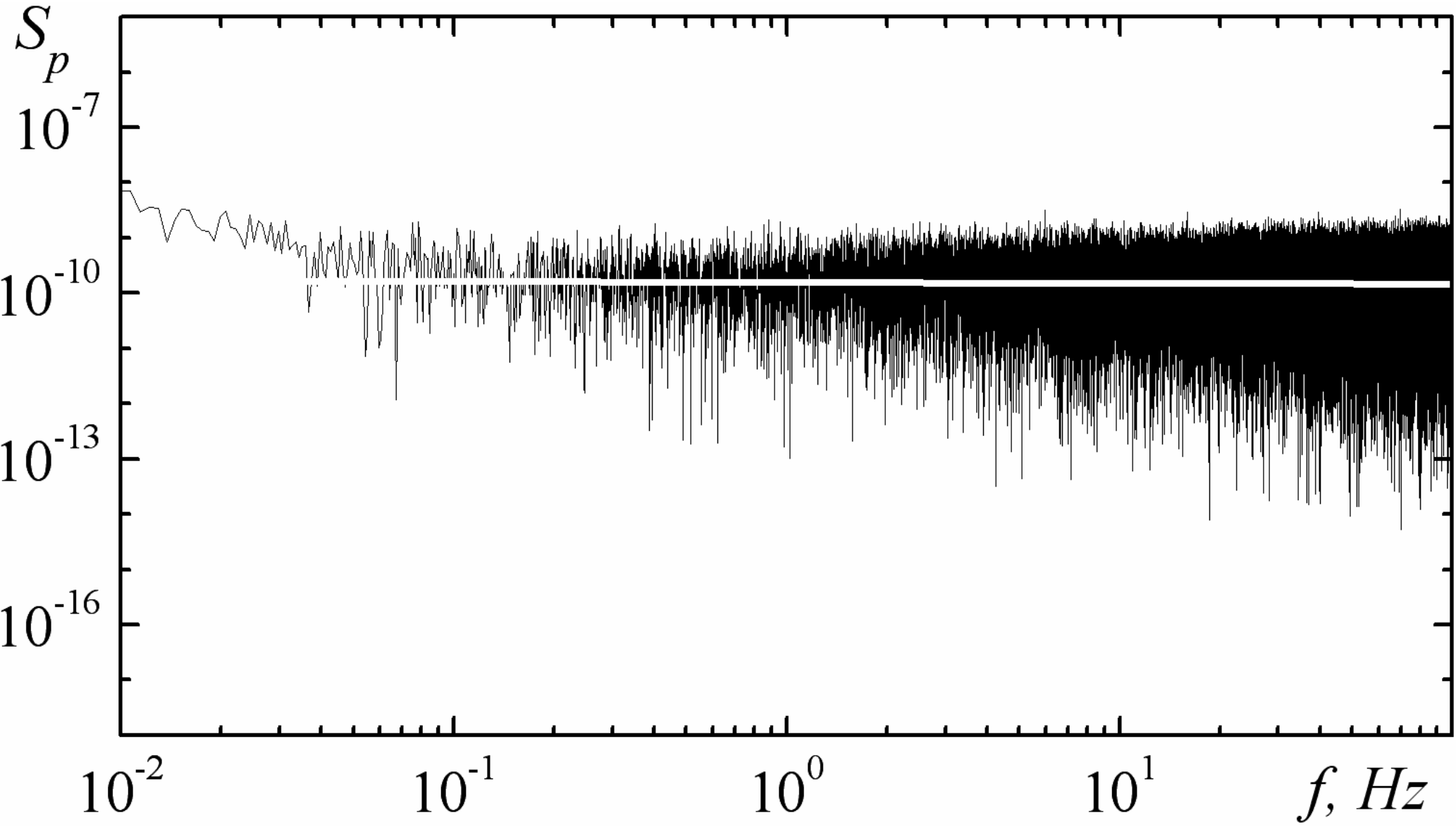}}}
\caption{Oscillation spectrum $S_p(f)$, related to shuffled series described by  fig.~\ref{fig6}.
White line corresponds to relationship $S_p(f)\propto 1/f^{0.017}$. Power $S_p$ is measured in conventional units.}
\label{fig9}
\end{figure}

According to above mentioned, we can conclude that in this case multifractality is caused by power-law distribution function. To detect different time
correlations which may be present in the system, we need to shuffle series and then, again calculate multifractal characteristics. In
figure~\ref{fig9} spectrum of the stress oscillations related to shuffled series, described by fig.~\ref{fig6}, is shown.
White line is the approximation of spectrum and can be described by relationship $S_p(f)\propto 1/f^{0.017}$, i.e., power almost does not depend on
frequency. It means that correlations in system are disappear. Given spectrum is related to white noise, $S_p(f)=\rm const$. Thus, shuffling
of the series causes elimination of correlations. But, since while shuffling the time series, neither addition nor substraction of the
series members are performed, distribution function stays the same.

Set of curves~2 in fig.~\ref{fig8} corresponds to shuffled series while set 1 is related to original ones. As we see $h(q)$ is a
straight line $h=0.5$, the spectrum function $f(\alpha)$ presents a narrow peak with small width,
and $\tau(q)$ is a straight line with constant slope. Denoted peculiarities are related to monofractal system, and value  $h=0.5$
corresponds to uncorrelated series. Thus, for considered system multifractality is caused by power-law distribution
function and by different correlations. If power-law dependence of $P_a(\sigma)$ is broken, or correlations vanish, the
multifractality is eliminated.

\section{Conclusion}

Using the homogenous rheological model ultrathin lubricant film melting has been investigated. Basic parameters are shear stress and
strain, as well as temperature of lubricant. Four regimes of lubricant behavior, characterized by different sets of maximums
of stresses distribution function, have been found. Stress time series have been obtained for each regime by numerical modeling of the
Langevin equation, and it has been shown that at specific parameters time series are multifractal. All basic multifractal characteristics have been
calculated, and it has been shown that multifractality is caused by different time correlation and also by power-law distribution in the limited
range of the stress values. When temperature noise intensity is much larger than intensities of stress and strain noises a power-law distribution
can be observed. According to above examination, multifractal time series are realized only for the dry friction ($\it DF$)
and $stick-slip$ ($\it SS$) domains, since only for this regimes power-law distribution is observed.

\section{Acknowledgments}
We express our gratitude to Dr. A. S. Kornyushchenko for attentive reading and correction of the manuscript. We glad to thank the State 
fund of fundamental researches of Ukraine (grants $\Phi$25/668-2007, $\Phi$25/97-2008) for support of the work.

\end{document}